\documentclass[journal=jacsat,manuscript=article]{achemso}
\usepackage{xcolor}
\usepackage{graphicx}
\usepackage{graphics}
\usepackage{scalefnt}
\usepackage{adjustbox}
\usepackage{float}

\usepackage{textcomp}

\usepackage[version=3]{mhchem} 

\author{Elizane E. de Moraes}
\affiliation{Instituto de Física, Universidade Federal da Bahia, Campus Universitário de Ondina, Salvador 40210-340, BA, Brazil}

\email{elizane.fisica@gmail.com}

\author{João Victor Lemos Vale}
\affiliation{Instituto de Física, Universidade Federal da Bahia, Campus Universitário de Ondina, Salvador 40210-340, BA, Brazil}

\author{Bruno H. S. Mendon\c{c}a}
\affiliation{Departamento de Física, ICEX, Universidade Federal de Minas Gerais, CP 702, 30123-970 Belo Horizonte, Minas Gerais, Brazil}

\author{Ernane de Freitas Martins}
\affiliation{Catalan Institute of Nanoscience and Nanotechnology - ICN2 (CSIC and BIST), Bellaterra, 8193 , Barcelona, Spain}

\author{Helio Chacham}
\affiliation{Departamento de Física, ICEX, Universidade Federal de Minas Gerais, CP 702, 30123-970 Belo Horizonte, Minas Gerais, Brazil}

\author{Pablo Ordej\'on}
\affiliation{Catalan Institute of Nanoscience and Nanotechnology - ICN2 (CSIC and BIST), Bellaterra, 8193 , Barcelona, Spain}

\title[An \textsf{achemso} demo]
  {Water Flow Through Polar and Non-Polar
Nanopores: Insights from Multiscale Simulations}

\abbreviations{IR,NMR,UV}

\begin{document}

\begin{abstract}
Global water stress has emerged as a critical challenge, driving the search for advanced membrane materials that enable efficient, selective water filtration and transport. In this context, two-dimensional nanoporous membranes provide an ideal platform to elucidate how atomic-scale structure and electronic polarization govern water flow under extreme confinement. In this study, we employ multiscale simulations to investigate the effect of water flow through nanopores in graphene and hexagonal boron nitride (hBN) membranes. Our results reveal significantly higher water flow in hBN membranes than in graphene. This enhanced flow is attributed to the asymmetry of the hBN pores, which induces an electric dipole moment, as confirmed by quantum-mechanical (QM) calculations. Classical molecular dynamics simulations further demonstrate that water molecules exhibit a random distribution with no preferential orientation near the graphene pores, whereas hBN induces strong structuring. Furthermore, hybrid quantum mechanics/molecular mechanics (QM/MM) simulations indicate that the dipole moment of the hBN pore increases in the presence of water, as evidenced by the average charge distribution. Conversely, the symmetric nature of graphene pores results in non-polar characteristics, as verified by both QM/MM and QM calculations. These findings provide valuable insights into the distinct water-transport properties when flowing through graphene and hBN nanopores, with potential implications for designing advanced nanofiltration membranes.
\end{abstract}


\section{Introduction}\label{sec1}

Access to clean water is one of the most pressing challenges of the 21st century. Climate change, population growth, and industrialization are intensifying global water scarcity, demanding innovative solutions for efficient filtration and purification. Among emerging technologies, nanofluidics has revealed that water confined within nanostructures can exhibit transport behaviors that defy classical expectations, opening new avenues for membrane design \cite{lang1982anomalies,kell1975density,netz2001static,majumder2005enhanced,whitby2008enhanced,bagchi2013anomalies,holt2006fast,mendoncca2023water,valle2024accuracy}.

Since the pioneering work by Quin et al. \cite{qin2011measurement}, which showed that water can flow through carbon nanotubes up to 900 times faster than predicted by classical fluid dynamics, researchers have increasingly turned to the nanoscale to explore and manipulate water transport for advanced filtration technologies \cite{@10.1038/43844a,@10.1126/science.1126298,@10.1021/nl200843g,@nature/35102535,@10.1021/jp402141f, @10.1021/acs.jpcb.3c02889,@10.1166/jctn.2014.3488,@10.1021/jp709845u,dick2026carbon,wang2025spontaneous,wang2025interfaces,chu2025molecular,wu2025continuum,andreeva2021,carrion2025,liu2025effect,liu2025facilitating}. 

Advances in nanofluidics have demonstrated that water confined within nanostructures can exhibit anomalously high flow rates, selective permeability, and distinct molecular interactions with membrane surfaces \cite{@10.1021/acs.jpcb.3c02889,eijkel2005nanofluidics,sam2019water,trushin2025structure}. These transport phenomena are governed by a complex interplay of factors, including pore geometry, surface functionalization, interaction of water with the surface, and the presence of structural defects. Recently, the exploration of nanoporous two-dimensional (2D) materials, which exhibit unique structural and chemical properties that can be harnessed to create highly efficient, selective membranes for desalination, has attracted significant interest \cite{cao2025spatially,li2025fundamental,meskher2025recent,sudalaimuthu2025impact,liang2025role,wu2025next,peng2026nature,manikandan2025massively,lu2025nano,dronadula2026nonlinear}. 

Graphene (GR) and hexagonal boron nitride (hBN) are two prominent 2D materials under investigation for nanofiltration. While graphene is known for its mechanical strength and chemical stability, hBN offers distinct advantages in ion selectivity due to its polar nature since boron and nitrogen atoms carry partial positive and negative charges, which can be exploited to engineer pore terminations.  Additionally, the ability to functionalize 2D materials with various chemical groups enables tuning of hydrophilicity or hydrophobicity, further influencing water transport characteristics \cite{bakshi2021structure,xing2024advances,vatanpour2021comprehensive,nam2022review,zhou2024interlink,vafa2023comparative}. 
 However, challenges remain, particularly in scaling up membrane fabrication and maintaining pore stability. For instance, a recent experimental study by Dai et al. demonstrated that pores formed in hBN layers are unstable when exposed to air, undergoing morphological changes and acquiring a rounded shape \cite{nanopores_evol}.

Despite extensive studies on water flow in 2D membranes, the role of pore asymmetry and electronic polarization in enhancing water transport remains poorly understood. In particular, the interplay between molecular orientation, hydrogen bonding, and charge distribution at the pore interface has not been fully explored. Advances in computational modeling, particularly in multiscale approaches such as hybrid quantum mechanics/molecular mechanics (QM/MM) simulations \cite{bakowies1996hybrid,lin2007qm,crespo2003dft} can provide valuable insights into the atomic-level mechanisms governing water transport through these membranes. Complementary experimental techniques can enable precise measurements of permeability and selectivity, validating theoretical predictions and guiding material design.

In this study, we employ multiscale simulations, including classical molecular dynamics (MD), density functional theory (DFT), and hybrid QM/MM methods, to investigate water flow through graphene and hBN nanopores. We show that hBN exhibits significantly higher water flux than graphene, driven by pore-induced dipole formation and water-enhanced polarization. Our findings provide atomic-level insights into the mechanisms governing water transport in 2D membranes and offer guidance for the design of next-generation nanofiltration technologies.

\section{Methodology}

\subsection{Classical Molecular Dynamics}

To investigate the flux of water molecules across nanoporous membranes, we designed the simulation box as illustrated in Figure~\ref{system}. The cell dimensions are $L_x \approx 37 $ \AA, $L_y \approx 37 $ \AA~and $L_z = 70 $ \AA. Periodic boundary conditions were applied in all directions. The system contains two reservoirs filled with 1000 water molecules each, two graphene sheets (piston 1 and piston 2) located at the edges in the z-direction, and a 2D membrane drilled with a nanopore with variable size in its center located at the origin of the box. Both graphene and hBN membranes were constructed with the same crystallographic orientation, with the armchair direction aligned along the $x$-axis in the membrane plane ($xy$-plane), ensuring direct structural comparison between the two materials. 
Nanopores were constructed from an effective radius ($r_{\mathrm{eff}}$) chosen to produce effective diameters close to those of the armchair carbon nanotubes (7,7), (9,9), and (12,12), namely 0.95, 1.22, and 1.63 nm, respectively. However, since the pore is generated by removing atoms from a discrete crystal lattice, the resulting pore boundary does not exactly correspond to an ideal continuous circle. Consequently, the actual geometric diameter of the pore may differ slightly from the adopted effective diameter. The real pore diameter was estimated from the atomic positions at the pore edge, using the radial distance from the pore center, 

\begin{equation}
   r = \sqrt{x^2 + y^2},
\end{equation}
with $x$ and $y$ being the in-plane atomic coordinates relative to the pore center. The corresponding pore area was then calculated as
\begin{equation}
    A = \pi r^2.
\end{equation}
For graphene, the diameters obtained are 0.98, 1.18, 1.64 nm and, for hBN, 1.01, 1.28, 1.74 nm. Owing to the distinct lattice constants (2.46 \AA~for graphene and 2.50 \AA~for hBN). The membranes and pistons were treated as rigid structures, with nonbonded interactions described by a Lennard-Jones (LJ) potential for graphene, while both Lennard-Jones and Coulombic interactions were considered for hBN. The water molecules are represented by the TIP4P-2005 force field~\cite{Abascal2005-iu}. In this water model, there is only one LJ interaction site placed at the oxygen position, with parameters $\sigma_{OO}$ and $\epsilon_{OO}$. The positive charges $q_H$ are placed at the hydrogen positions, while the negative charge $q_M$ is placed at a distance $d_{OM}$ from the oxygen along the  H-O-H bisector. The cross-interaction parameters between all species are provided by the Lorentz-Beherlort mixing rules. All the force field parameters are summarized in Table~\ref{tab_parameters}.

\begin{table}[h]
	\begin{center}
	\caption{Force field parameters for the solid frameworks (GR and hBN) and water molecules. The units of the LJ parameters are $\mathrm{kcal.mol^{-1}}$ for $\epsilon_{XX}$ and \AA~for $\sigma_{XX}$, where $X=\mathrm{(C,O,H,B,N)}$. The charges are given in elementary charge (\textit{e}) units.}
		\begin{tabular}{ c c c c c c c }
			\hline
			   $\mathrm{H_2O}^{(a)}$ && $\mathrm{hBN}^{(b)}$ && $\mathrm{Gr}^{(c)}$ (Pistons) &  \\ \hline
              $\epsilon_{OO}$= 0.1852  && $\epsilon_{BB}$= 0.09491 && $\epsilon_{CC}$= 0.086 & \\
              $\epsilon_{HH}$= 0.0000  && $\epsilon_{NN}$= 0.14480 && $\sigma_{CC}$= 3.40   & \\
              $\sigma_{OO}$= 3.1589    && $\sigma_{BB}$= 3.45600   && $q_C$= 0.00  & \\
              $\sigma_{HH}$= 0.0000    && $\sigma_{NN}$= 3.36500   &&              & \\
              $q_M$= -1.1128           && $q_B$= 0.52400           &&              & \\
              $q_H$= 0.5564            && $q_N$= -0.52400        &&              & \\
              $r_{OH}$ (\AA)= 0.9572 &&                &&              & \\
              $\theta_{HOH}$ ($\mathrm{deg}$)= 104.52 &&           &&              & \\
              $d_{OM}$ (\AA)= 0.1546  &&               &&              & \\
            \hline 
		\end{tabular}
        \\
        \footnotesize{(a): Vega and Abascal~\cite{Abascal2005-iu}; (b): Ghoufi et al.~\cite{hBN_parameters}; (c): Hummer et al.~\cite{Hummer2001}.}
		\label{tab_parameters}
	\end{center}
\end{table}

\subsection{Simulation Protocol}

\begin{figure}
\centering
\includegraphics[width=1.0\textwidth]{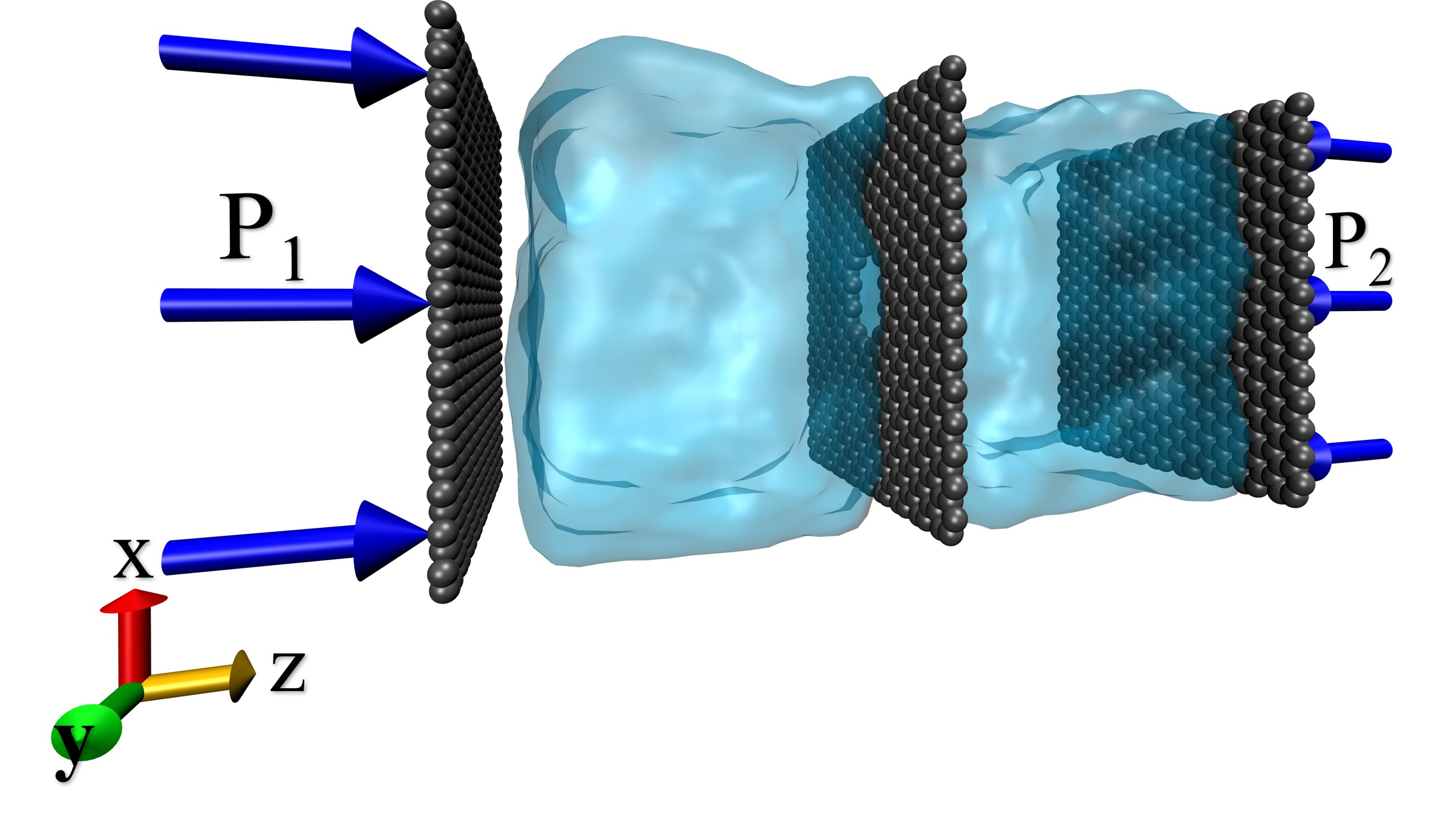}
    \caption{Perspective view of the simulation box, where the cyan transparent surface represents the water molecules, gray spheres represent the piston plates, and the central gray membrane with a pore represents the nanoporous 2D membrane located at the origin of the coordinate system.}
    \label{system}
\end{figure}

The simulations were performed by using the Large-scale Atomic/Molecular Massively Parallel Simulator (LAMMPS) software, with a timestep of 1.0 fs. The SHAKE algorithm was employed to maintain the membrane's and pistons' bond lengths and angles constrained. Coulombic long-range interactions were computed by the PPPM solver with a precision of $10^{-4}$. We used a cutoff of 12 \AA~for both LJ and Coulombic interactions. The Nosé-Hoover thermostat (NVT) is used to control temperature, and the Nosé-Hoover chaim barostat (NPT) is used to control the pressure. 
In the protocol used to create the pores and equilibrate the systems, the nanopore is geometrically present, but effectively blocked by buffer atoms placed inside the pore to prevent water permeation before the system reaches full equilibrium. These buffer atoms are defined as fluorine-type atoms and act as a temporary barrier, ensuring that no flow occurs during the equilibration stages. These buffer atoms are not part of the physical membrane and are used solely for numerical and thermodynamic stabilization. Once the confined water reaches equilibrium density and temperature, the buffer atoms are removed from the simulation, thereby opening the nanopore. The simulation protocol for each system is given as follows:

\begin{enumerate}
    \item Pre-equilibration in the NVE ensemble via a 0.1 ns molecular dynamics run. During this stage, the pistons remain frozen (acting as fixed rigid walls); although the fluid exerts forces on the pistons, these forces are not integrated, ensuring that the piston atoms undergo no displacement.
  \item Equilibration in the NVT ensemble at 300 K during 0.2 ns.
    \item Forces are then applied in the pistons to impose 1 bar to reach the water equilibrated densities at 300 K using the NPT ensemble for a 1.0 ns MD run. 
    \item Pistons are frozen in the new equilibrium position, and another NVT equilibration at 300 K is performed for 10 ns. 
    \item Nanopores are opened by removing the buffer atoms to enable water transport throigh the pore, and different forces are applied in each piston to mimic the pressure differences using the NPT ensemble during 10 ns at 300 K and different feed pressures.
\end{enumerate}
The water flows along the z-axis, normal to the membrane, driven by the pressure difference between both sides of the membrane. The pressure difference is caused by pistons 1 and 2, as illustrated in \ref{system}, being introduced by applying external forces F on each atom of the pistons in the z-direction.

\begin{equation}
    F = \frac{P \cdot A}{n}~,
    \label{eq1}
\end{equation}


\noindent where $n$ is the number of atoms, $A$ is the surface area, and $P$ is the pressure on the surface. Piston 2 (right) has a fixed pressure of 0.1 MPa, while piston 1 (left) was set to achieve left-right pressure differences 
of 100 MPa, 200 MPa, 300 MPa, 400 MPa, and 500 MPa. 


\begin{figure}
\centering
\includegraphics[width=0.9\textwidth]{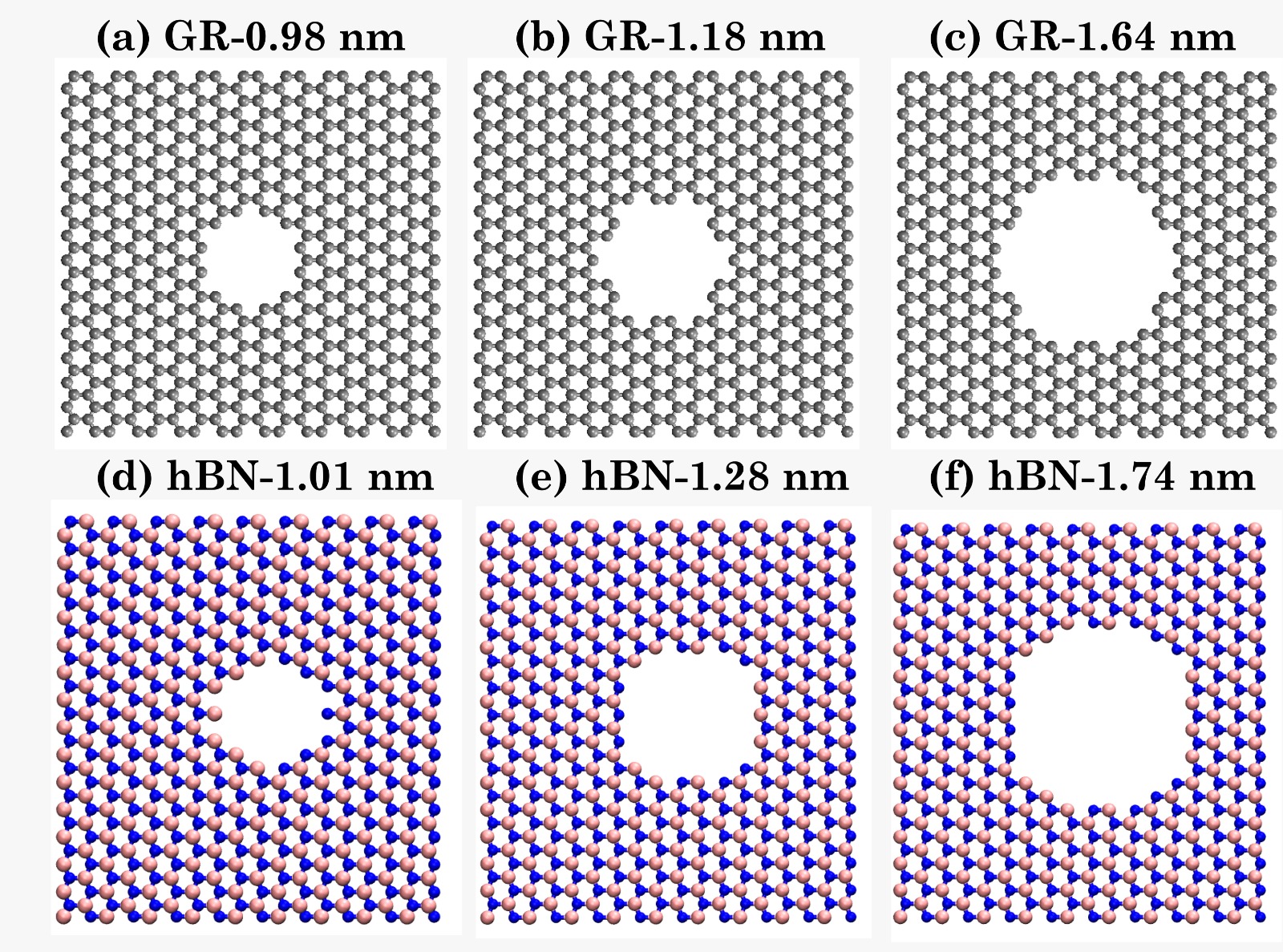}
    \caption{Nanoporous membrane structures with different pore diameters. (a)–(c) represent the GR membrane and (d)–(f) represent the hBN membrane. The numbers shown at the top of each panel indicate the nanopore diameters (in nm). Carbon atoms are shown in gray, boron atoms in blue, and nitrogen atoms in pink.}
    \label{fig:poro}
\end{figure}


We study the dynamics of the liquid, considering the flow rate of water molecules through the nanopore:

\begin{equation}
    \phi_{H_2O} = \lambda \; V_{mol} \; v~,
    \label{eqflow}
\end{equation}
where $\lambda (\mathrm{molecules}.\mu m^{-1}$) is the linear number density of water molecules, $V_{mol} (\mu m^3.molecule^{-1})$ is the average volume of a water molecule and $v (\mu m.s^{-1}$) is the water flow velocity which is acquired from the least-square linear regression line fitted to the data cloud which relates the average molecular displacement as a function of the time, as taken from the MD trajectory file. The calculations are performed inside a rectangular box centered at the pore, and with dimensions spanning the whole supercell in surface directions, and $\Delta z = 4$\AA~ along the direction of the flow.

\subsection{QM/MM Calculations}

Our calculations have been performed within the framework of DFT  \cite{PhysRev.140.A1133} as implemented in the SIESTA program\cite{soler2002siesta, sanchez1997density,garcia2020siesta},  including its module to perform hybrid QM/MM simulations \cite{crespo2003dft,Sanz2011,Feliciano2015,Feliciano2018,i2026lennard}. We split the system into a QM (membrane) and a MM (water) part, using snapshots from classical MD simulations. The pistons are removed for the QM/MM runs. We used the well-known exchange–correlation DRSLL functional \cite{dion2004van} for the QM part, with norm-conserving pseudopotentials in the Kleinman-Bylander factorized form \cite{testeKB,martins91}, and a double-zeta plus polarization (DZP) basis set for C, B, and N atoms. A real-space grid was used with a mesh cutoff of 250 Ry, and the Brillouin zone was sampled using a k-grid cutoff of 20 \AA. The Coulombic potential generated by the MM atoms for each snapshot is considered using the SPC water model \cite{wu2006flexible}, for simplicity purposes, as well as its LJ parameters for the oxygen atoms to account for dispersion interactions between QM and MM atoms.

\section{Results}\label{sec3}
\subsection{Classical Molecular Dynamics}

\begin{figure}[h]
    \centering
    \includegraphics[width=1.0\textwidth]{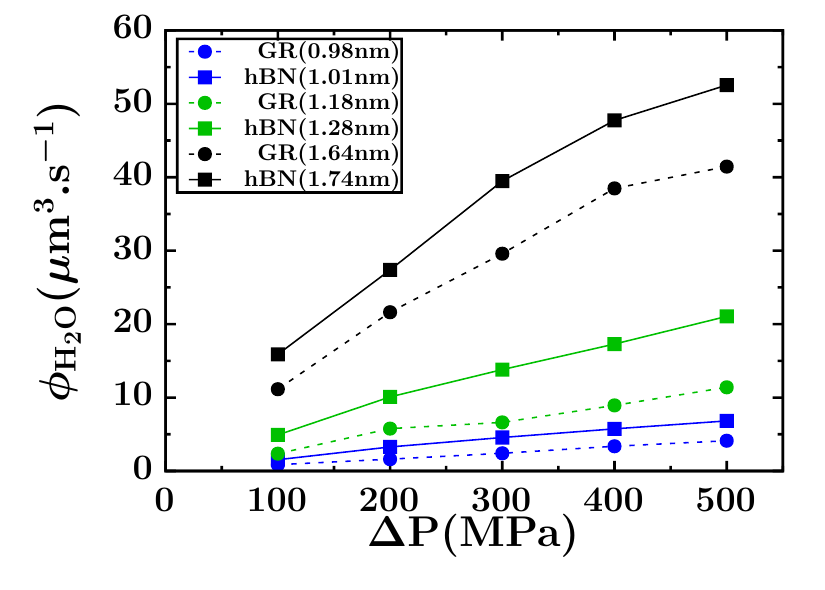}
    \caption{Water flow rates through graphene (circles) and hexagonal boron nitride (squares) nanopores as a function of pore diameter: 0.98, 1.18, and 1.64 nm for graphene, and 1.01, 1.28, and 1.74 nm for hBN (blue, green, and black, respectively).}
    \label{fig:flow1}
\end{figure}

Figure~\ref{fig:flow1} illustrates the flow of water as a function of the applied pressure for GR and hBN nanopores with varying pore diameters, computed according to Equation \ref{eqflow}. For each system, the flux is roughly linear with the applied pressure difference, although substantial deviations from linearity are observed for the largest pores, specially for large pressures.
For a given pressure and membrane material, flux increases with diameter because the greater pore area provides more pathways for water molecules to permeate. Most importantly, hBN membranes exhibit  a larger water flux than graphene membranes across all applied pressure differences and pore diameters. This finding aligns with the observations of Garnier et al.~\cite{garnier}, who attribute the enhanced permeability of nanoporous hBN to a reduction in the surface tension of water molecules in the normal direction.

To gain deeper insights into the molecular flow mechanisms related to the diameter of the pores in different materials, we generated oxygen density maps on the surface of the nanopores, as shown in Figure~\ref{fig:maps} at a pressure of 100 MPa (analogous maps at other pressures can be found in the supplementary information - SI).  The density maps represent the normalized spatial distribution of oxygen atoms within the nanopore region. To construct these maps, a rectangular domain with dimensions 
$ d_x = L_x $, $ d_y = L_y $, and $ d_z = 4$ \AA~ was selected, centered at the pore. 
This domain was subdivided into smaller subdomains (bins) with dimensions 
$ d_x = L_x/N $ $d_y = L_y/N$ \AA~ and $ d_z = 4$ \AA~, where 
N is the number of partitions along the x and y directions. For each subdomain, the local oxygen atom density was calculated at every time step  from the MD trajectory file. A time-average density was then computed for each bin, yielding the final two-dimensional density map. For visualization, the density values in each map were normalized by the maximum density value of that same map, resulting in values between 0 and 1, which were then multiplied by 100 and expressed as normalized density on a percentage scale. The program used to generate these maps was adapted from \cite{vmd_program}.

The density map shows that water molecules are preferentially distributed near the pore walls for all pores considered. For the smallest pore in GR shown in Figure~\ref{fig:maps} (a) the distribution along the pore edge is circular and quite uniform. For the GR pore in Figure~\ref{fig:maps} (b), water molecules adopt a hexagonal structure along the pore wall boundary (color red), organizing themselves into six localized and well-defined density maxima. These density maxima represent preferential interaction sites where water molecules spend most of their time due to interaction with the GR pore edge.  
The pore in this case is large enough to allow for some water molecules to pass through the center of the pore and of the ring of water molecules in contact with the pore edges, as seen by the light region in Figure~\ref{fig:maps} b. For the largest pore diameter shown in Figure~\ref{fig:maps} (c),  a hexagonal structure forms, following the contour of the nanopore,  while a second circular layer clearly develops at the pore center. The distance between both layers is approximately  $2.5$-$3$ \AA~(see Figure S3  in the supplementary material), which is a characteristic distance between layers in water in contact with graphene surfaces \cite{doi:10.1021/acsomega.7b00365} and nanoconfined water \cite{CALERO2020114027}.

\begin{figure}[H]
    \centering
    \includegraphics[width=0.9\textwidth]{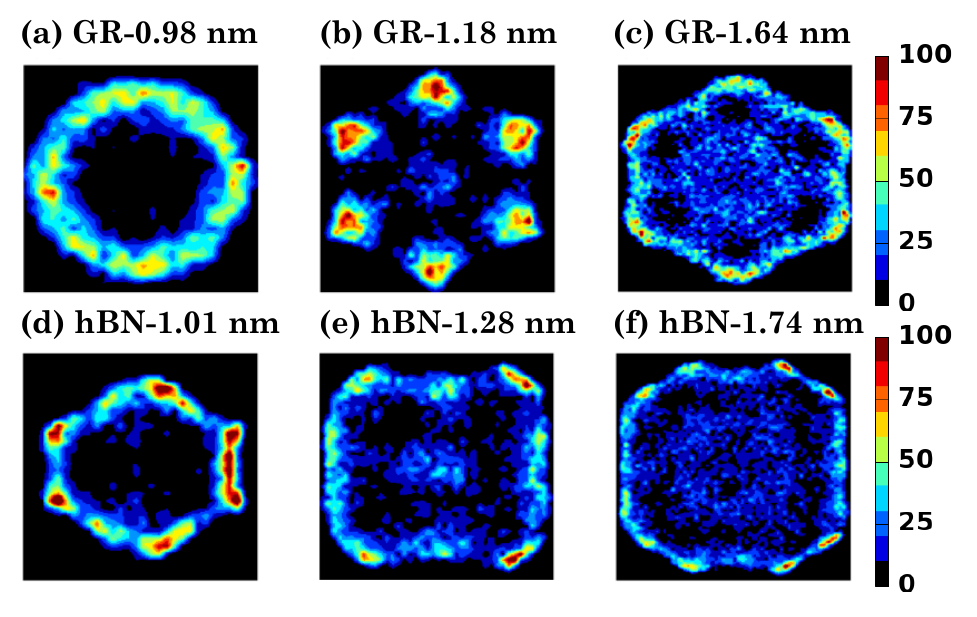}
    \caption{Oxygen density maps for membranes under a pressure difference of $\Delta P = 100\mathrm{MPa}$. Red regions indicate a high probability of water molecule presence, blue regions indicate low probability, and black regions correspond to areas where no water molecules were detected.}
    \label{fig:maps}
\end{figure}

As for the hBN pores, the distribution of water molecules follows the general trends as in the case of graphene, with a single layer of molecules at the edge of the pore for the smallest diameter, the appearance of a central row of molecules at the intermediate one, and the development of a second circular central layer for the largest diameter. The details of the distribution of molecules is, nevertheless different from that of graphene. The the narrowest pore, the arrangement of molecules is hexagonal, following closely the contour of the pore, in contrast with the case in graphene, where the shape was circular. This indicates a stronger interaction between the water molecules and the pore edges for the hBN case, which was to be expected as the N and B atoms have partial negative and positive charges, respectively, which interact through Coulomb interactions with the charges of oxygen (negative) and hydrogen (positive) in water. Some degree of asymmetry is also observed in the distribution of oxygen atoms for the three pore sizes for hBN, again in contrast with the case of graphene. This asymmetry is due to the fact that the pores have walls with facets where either N or B is exposed, leading to asymmetric interactions and molecular orientations and distributions.

Before proceeding, it is important to highlight the intrinsic net dipole moment present at hBN pores, a feature that GR pores do not share. In classical MD simulations, partial charges are assigned to each atom; thus, the opposite-sign charges of B and N naturally establish an electric field across the pore. This local field can influence water
configuration, orientation and transport through the membrane. A clearer way to analyse this effect is by examining the orientation of water molecules residing at the pore, particularly the angular distribution of their dipoles relative to the pore axis.

\begin{figure}[H]
    \centering
    \includegraphics[width=1.0\textwidth]{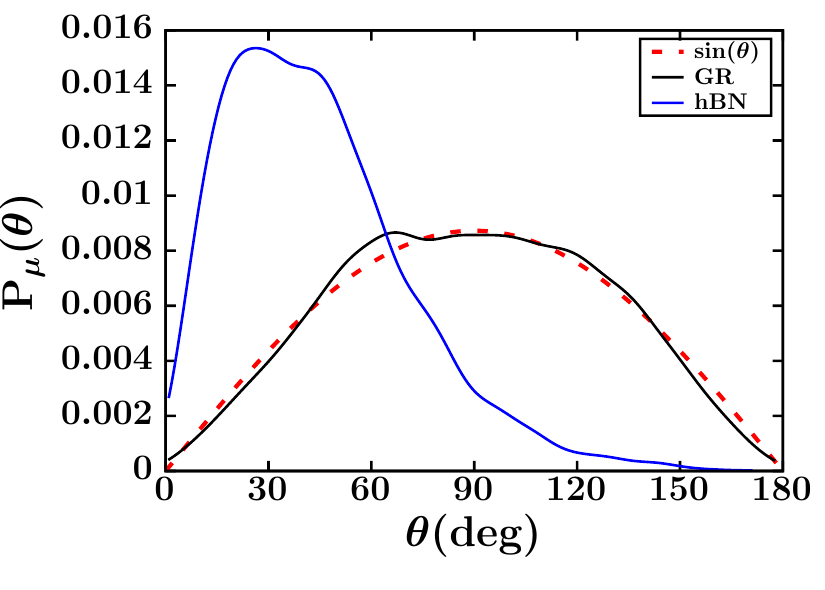}
    \caption{Analysis of the water dipole angle distribution along the x-direction for the smallest graphene (0.98 nm) and hBN (1.01 nm) nanopores under $\Delta P = 100\mathrm{MPa}$. The vertical axis of the plot is the x component of the sum of the electric dipole
moments of the water molecules located within the pore region.    }
    \label{fig:dipole}
\end{figure}

To study the water orientation residing at the pores, we analyzed the angular distribution of the water dipole moment vectors relative to the axis defined by the pore dipole (see Figure~\ref{fig:dipole}).

For the case of the GR nanopores, the distribution of the dipole angle follows quite closely the random distribution (proportional to the sine of the angle). This shows that, for water passing through the nanopores in GR, there are no preferential directions for their dipoles, which is a consequence of the lack of electrostatic interactions between the molecules and the pore edges. For hBN, on the contrary, the distribution deviates strongly from the random case:  it shows a strong asymmetry, with a preference towards the water dipoles pointing towards the positive $x$-axis. This occurs because these dipoles tend to align with the electric field generated by the negatively charged nitrogen atoms and the positively charged boron atoms present on each side of the nanopore edges: the hydrogen atoms preferentially point toward the negatively charged side, while the oxygens are attracted towards the positive side. 
As pointed out by Gao et al.~\cite{Gao2017}, this type of interaction is crucial for water transport and constitutes a dominant effect when compared to pores terminated solely by boron atoms with similar diameters.

\begin{figure}[H]
  \centering
    \includegraphics[width=1.0\textwidth]{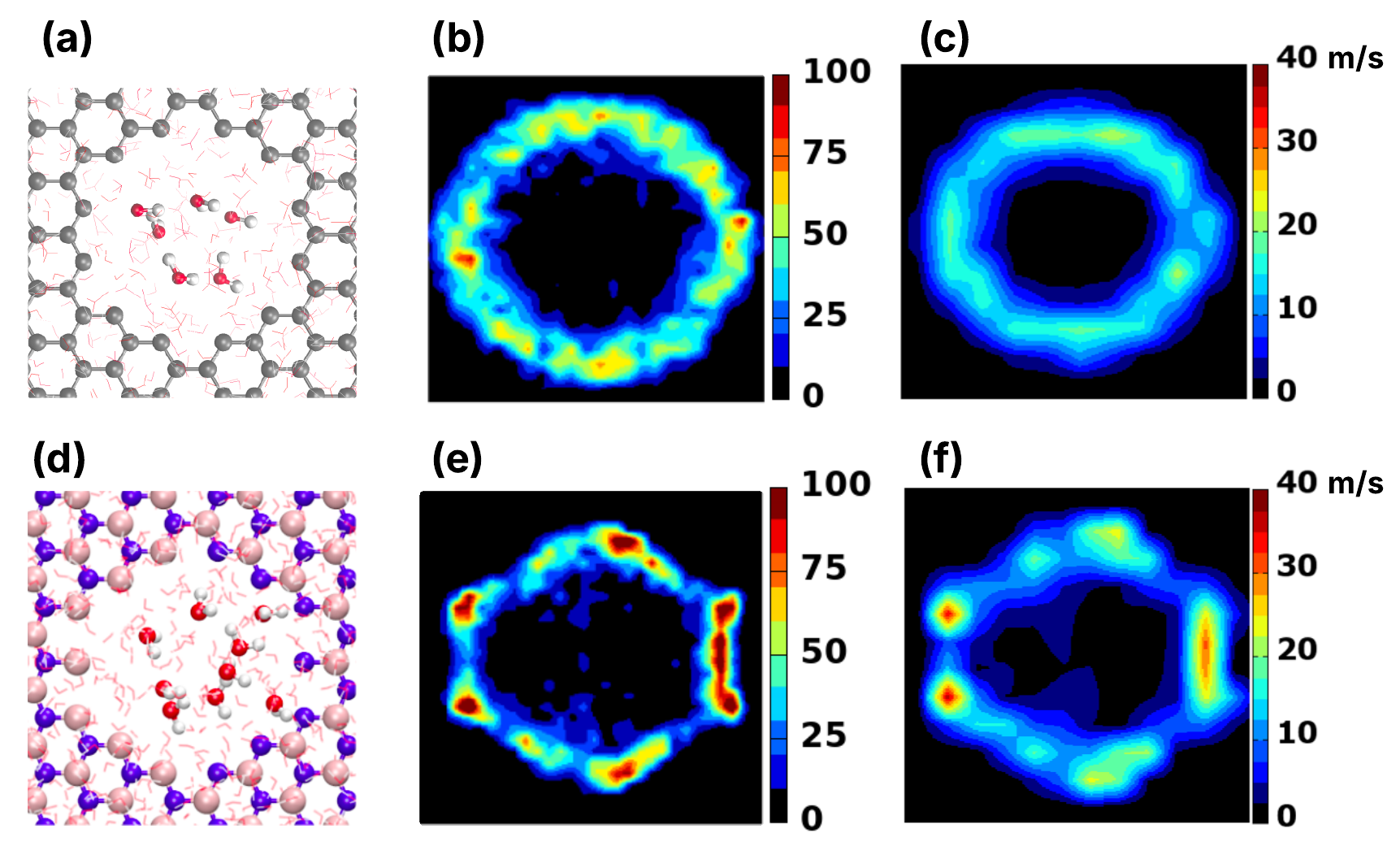}
        \caption{Representative snapshot of water on the pore region (a,d), oxygen density maps (b,e), and axial velocity profiles of water molecules along the z-direction (c,f) for membranes under a pressure difference of $\Delta P = 100\mathrm{MPa}$ for graphene (upper panel) and hexagonal boron nitride (lower panel).}
  \label{fig:velocity}
\end{figure}

To understand the mechanisms responsible for the difference in water flux between GR and hBN nanopores shown in Figure~\ref{fig:flow1}, we evaluated the axial velocity profiles of water molecules passing through the smallest pore  under an applied pressure difference of 100 MPa, as shown in Figures~\ref{fig:velocity}(c) and (f) together with the oxygen maps already presented in Figures~\ref{fig:maps}(a) and (d). The axial velocity profiles of water were computed by selecting a region of interest in the shape of a parallelepiped with dimensions $(dx = L_x)$, $(dy = L_y)$, and $(dz = 4)$~\AA, where the thickness along the $z$ direction is the same as that adopted in the flux calculation. This region was then subdivided into smaller bins with dimensions $(dx = L_x/N)$, $(dy = L_y/N)$, and $(dz = 4)$~\AA, where N is the number of partitions along each in-plane direction. For each bin and at each time step, the magnitude of the z-component of the velocity, $|v_z|$, was calculated for every oxygen atom within that bin, taking the oxygen atom position as an approximation to the molecular center of mass of the water molecule. These values were then averaged over all oxygen atoms present in the bin at that time step. This procedure was repeated for all time steps considered in the analysis. Finally, a time average of $|v_z|$ was computed for each bin, so that the resulting value already corresponds to an average both over the oxygen atoms within the bin and over the trajectory frames. The final velocity map is then constructed from an output file containing the (x, y) coordinates of each bin, along with the corresponding mean value of $|v_z|$. The velocity was converted to m/s in order to follow the convention adopted in the literature \cite{joseph2008carbon,heiranian2015water,suk2010water}.

For the GR membrane, water exhibits a symmetric behaviour, as observed in both the oxygen density and the average particle speed illustrated in Figures~\ref{fig:velocity}(b) and (c). For the hBN case, the velocity shows a strong asymmetry, with a tendency towards higher velocities for the water molecules closer to the N motifs of the pore edge (right side in Figure~\ref{fig:maps} (e, d), although it is also large at the corners exposed to three boron atoms, where maxima for both the oxygen concentration and velocity appear (left side of Figure~\ref{fig:maps} e, d. The alignent of the water dipoles, with O oriented towards B and H towards N, in conjunction with a H-bond network between neighbour water molecules, allows a high density. This indicates that, although the electrostatic interactions are sufficiently large to lead to a strong structure of the water passing through the pore, it is not large enough as to lock the molecules to the pore walls. On the countrary, flow is more facile in the hBN case than in GR, where the interactions are smaller and there is less structuring.

Finally, it is worth comparing panels b and e with c and f in Figure~\ref{fig:velocity}. For GR, the velocity and density maps exhibit an almost perfect spatial correspondence, indicating that regions with higher water density are precisely those where the local flow velocity is enhanced. A similar trend is observed for the hBN pore: the areas where water molecules accumulate (highlighted in red in the density maps) coincide with the regions of highest velocities in the corresponding velocity maps. This clear correlation reveals that water molecules preferentially traverse the pore through the B- and N-terminated edges, and moreover, they do so with significantly higher velocities along these pathways.

However, a notable deviation emerges at the B–N frontier, located at the central region of both maps in Figure~\ref{fig:velocity}. Although this area exhibits comparatively high water density, the corresponding velocities are substantially lower than those measured near the pure B or N edges. This apparent discrepancy can be rationalized by the stronger water–pore interactions present at the B–N interface: hydrogen atoms in the water molecules experience attractive interactions with N sites, whereas oxygen atoms experience analogous attractions with B sites. These competing interactions increase the residence time of molecules in this region, thereby reducing their effective velocity despite the elevated local density.
    
\subsection{QM and QM/MM calculations}

The analysis presented above demonstrates that water transport is significantly enhanced in hBN membranes compared to graphene. The angular distribution reveals a well-defined orientational ordering of water molecules inside hBN nanopores, driven by the intrinsic dipole associated with the asymmetric B–N pore termination.  However, the empirical model used in the previous calculations assumes certain fixed partial charges, which deserve a validation though accurate first principles calculations. Similarly, the passage of water molecules though the pore  could significantly affect the electronic structure of the membrane, and in turn the partial charges on its atoms and therefore the interaction with the water molecules.  To assess these two issues, we have performed simulations based on DFT for the membranes in vacuum, and QM/MM for the membranes in the presence of water. The results are presented in this section.

\begin{figure}[H]
    \includegraphics[width=1.0\textwidth]{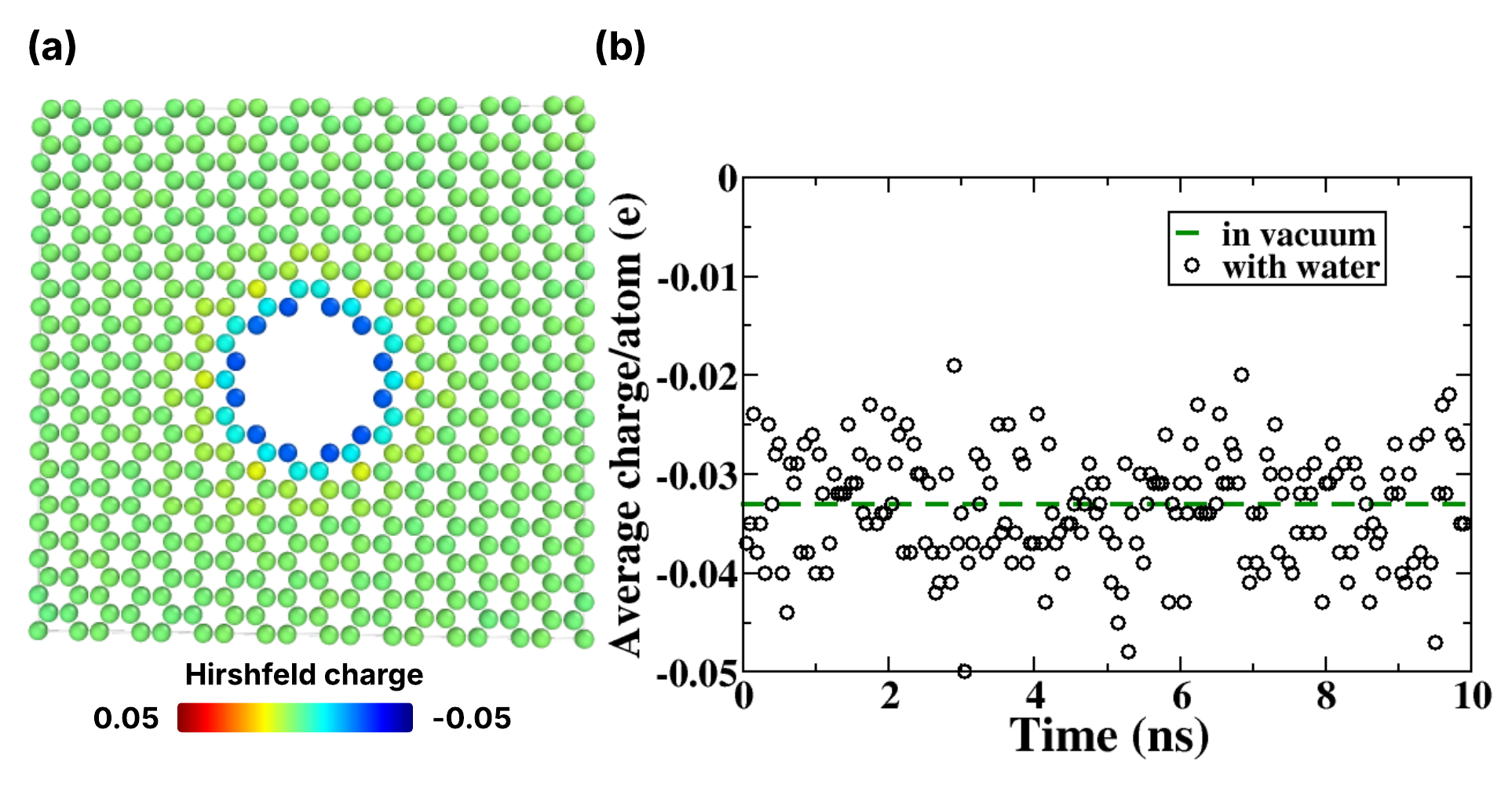}
    \caption{(a) Average atomic charges for the graphene membrane obtained from quantum mechanics/molecular mechanics (QM/MM) calculations over 200 snapshots extracted from the classical molecular dynamics. (b) Average charge per atom in the pore region showing the charges obtained in a vacuum (QM dotted line) and those obtained in the presence of the water molecules (QM/MM circles). 
    }
    \label{fig:QMMMGR}
\end{figure}

To assess the electronic structure of the membranes investigated and how the presence of water molecules in the pore region affects it, we extracted 200 snapshots from classical MD simulations to perform hybrid QM/MM calculations only for the smallest pore diameters. The GR and hBN membranes were treated at the QM level, while all water molecules were described at the MM level. Figure~\ref{fig:QMMMGR} presents the average atomic charges as well as the charge on the pore atoms as a function of time, obtained from all snapshots for the GR membrane. 

As shown in panel (a) of Figure~\ref{fig:QMMMGR}, carbon atoms located more than one bond away from the pore edge exhibit an average charge of around zero, indicating minimal influence by the MM water molecules. 

However, pore-edge carbon atoms already exhibit a net negative charge in vacuum because of missing bonds at the pore-edge atoms. Including water in QM/MM does not significantly change the mean edge charge, but introduces small temporal fluctuations around the vacuum value, indicating weak membrane–water electronic coupling in graphene. 
Panel (b) of Figure \ref{fig:QMMMGR} compares the average atomic charges in the pore region under two conditions: QM-only (dotted line), without water, and QM/MM (circles), with water included. The charge variation in GR is relatively small, and  water's influence on the electronic structure is modest, indicating weak coupling between the water molecules and the GR pore. 

\begin{figure}[H]
\includegraphics[width=1.0\textwidth]{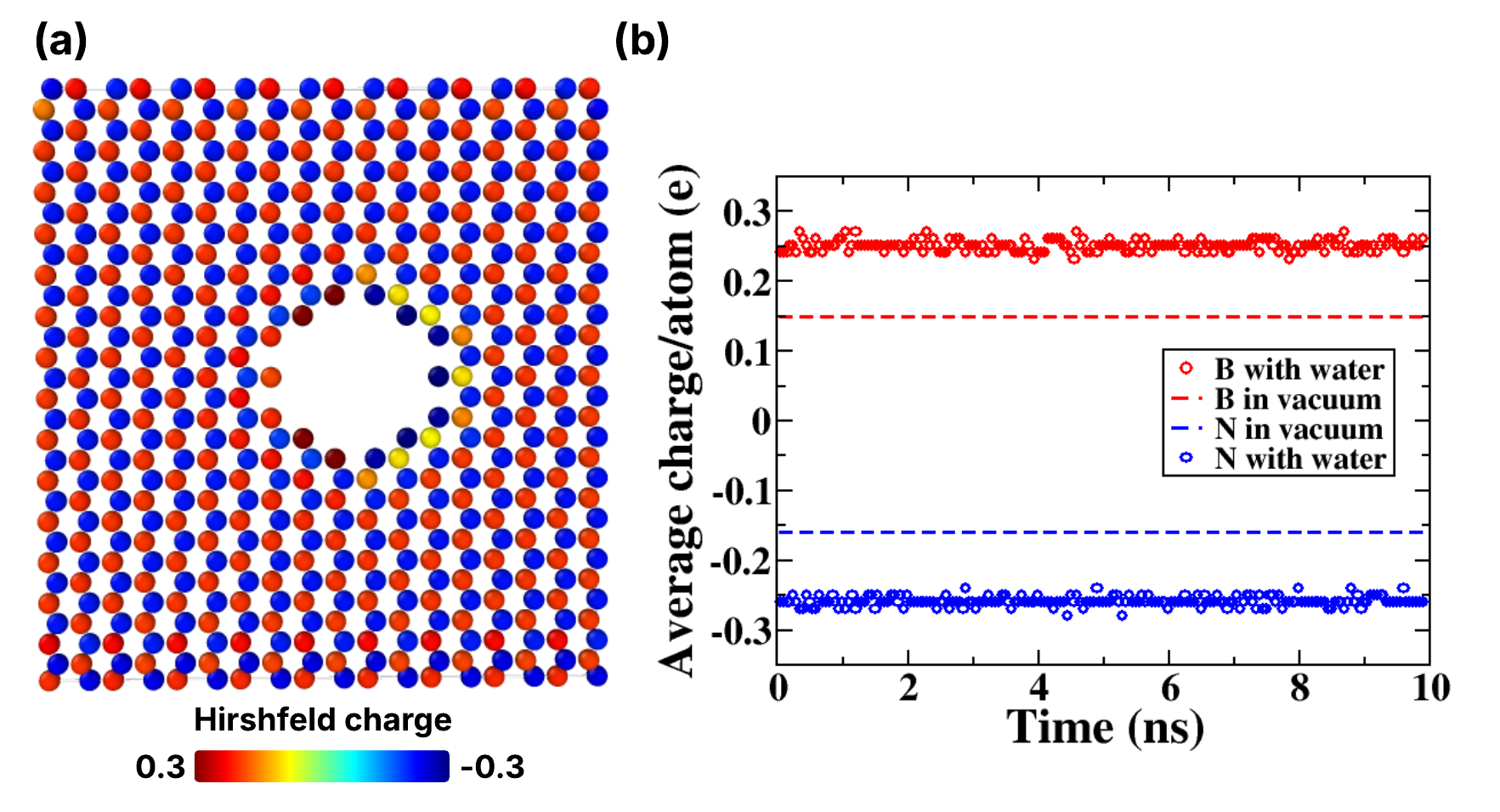}
    \caption{(a) Average atomic charges for the hexagonal boron nitride membrane obtained from quantum mechanics/molecular mechanics (QM/MM) calculations over 200 snapshots extracted from the classical molecular dynamics. (b) Average charge per atom in the pore region showing the charges obtained in a vacuum (QM dotted line) and those obtained in the presence of the water molecules (QM/MM circles). Blue color in (b) indicates the N-side, while red color indicates the B-side.
    }
    \label{fig:QMMMhBN}
\end{figure}

The analysis of the charge distribution for the hBN membrane is presented in Figure~\ref{fig:QMMMhBN}. As expected, boron and nitrogen atoms exhibit positive and negative charges, respectively, both in the bulk region and near the pore, reflecting the intrinsic polarity of the B–N bond. 
 
Figure \ref{fig:QMMMhBN} b represents the average charge per atom in the pore region, distinguishing between the B-side (red line) and N-side (blue line). The inclusion of water in the QM/MM calculations enhances charge separation relative to the vacuum case (dotted lines), indicating a significant water-induced polarization. Notably, the dipole formed across the pore by the intrinsic charge asymmetry between B and N atoms persists throughout the simulation, suggesting a stable electrostatic environment that may influence water transport and orientation.

The QM/MM results provide electronic-level insight consistent with the trends observed in classical MD simulations. While the classical force field assumes fixed atomic charges, the QM/MM calculations reveal that the magnitude of the pore dipole moment is enhanced in the presence of water.

While classical MD simulations revealed enhanced water flux and orientational ordering in hBN nanopores, QM/MM calculations clarify the electronic origin of this behavior. An intrinsic dipole moment, arising from the B–N charge asymmetry at the pore edge, is already present in vacuum and establishes a persistent electric field across the pore. The inclusion of water further increases the magnitude of this dipole moment, revealing solvent-induced amplification of the intrinsic polarization. This indicates that the fixed-charge description employed in classical MD likely underestimates the strength of electrostatic alignment effects.

In the GR pore, the charge fluctuations and the absence of a strong water-induced polarization are consistent with the lower flux and the nearly random orientational distribution of water molecules (Figure~\ref{fig:dipole}). The QM/MM calculations provide an electronic-level perspective that complements the classical simulations: they show weak membrane–water electronic coupling in GR and reveal solvent-amplified polarization in hBN, highlighting a limitation of the fixed-charge description used in classical MD.

\section{Conclusion}\label{sec4}

In this study, we investigated water transport through graphene (GR) and hexagonal boron nitride (hBN) nanopores using a multiscale simulation approach combining classical molecular dynamics (MD), density functional theory (DFT), and hybrid quantum mechanics/molecular mechanics (QM/MM) calculations. Classical MD simulations revealed that hBN membranes exhibit significantly higher water flux than GR, a behavior attributed to the structural asymmetry and electrostatic characteristics of the hBN pore. 

Water molecules near GR pores showed random orientation and symmetric distribution, whereas in hBN pores, a strong structuring and directional alignment of dipoles were observed. This alignment facilitates more organized hydrogen bonding, contributing to the observed increase in water flux. In addition, oxygen density maps and velocity profiles showed that water molecules near graphene present a more uniform but less efficient flow. These findings underscore the importance of pore geometry and atomic composition in modulating water behavior at the nanoscale.

QM and QM/MM calculations further elucidated the electronic mechanisms underlying these differences. While GR pores remained largely non-polar, with minimal charge redistribution, hBN pores developed a stable dipole due to the intrinsic charge asymmetry between boron and nitrogen atoms. This dipole moment was amplified in the presence of water, as shown by the average charge distribution in QM/MM simulations, indicating water-induced polarization. The resulting internal electric field in hBN pores promotes dipole alignment and directional water flow, contributing to the enhanced permeability observed in classical MD simulations.

These results demonstrate that the superior water transport properties of hBN nanopores arise from a combined effect of geometric confinement and water-induced electronic polarization. The integration of classical and quantum-level approaches offers a comprehensive understanding of nanoscale fluid transport and highlights the potential of hBN membranes for high-performance filtration and nanofluidic applications.\\
\\
\\
Supporting Information.\\

Additional analyses, including oxygen density maps at different pressure differences and radial density profiles, are provided in the Supplementary Material.

\section{Acknowledgement}

This work is funded by the Brazilian scientific agencies Fundação de Amparo à Pesquisa do Estado da Bahia (FAPESB), Fundação de Amparo à Pesquisa do Estado de Minas Gerais (FAPEMIG), Comissão de Aperfeiçoamento do Pessoal de Ensino Superior (CAPES), Conselho Nacional de Desenvolvimento Científico e Tecnológico (CNPq), and the Brazilian Institute of Science and Technology (INCT) in Carbon Nanomaterials, with collaboration and computational support from Universidade Federal da Bahia (UFBA),  Universidade Federal de Minas Gerais (UFMG.  In addition, the authors acknowledge the National Laboratory for Scientific Computing (LNCC/MCTI, Brazil) for providing HPC resources of the SDumont supercomputer, which have contributed to the research results reported within this paper. URL: http://sdumont.lncc.br. BHSM acknowledges FAPEMIG for the postdoctoral fellowship (Grant No. APD-01962-25)
The authors thankfully acknowledge the computer resources at MareNostrum and the technical support provided by Barcelona Supercomputing Center (QHS-2024-2-0035, QHS-2024-3-0022, and QHS-2025-1-0037). 
 EFM and PO were supported by Grant PID2022-139776NB-C62 from the Spanish MCIN/AEI/10.13039/501100011033. 
ICN2 is funded by the CERCA Program (Generalitat de Catalunya) and the Severo Ochoa Program from Spanish MINECO (Grant No. CEX2021-001214-S). This project has received funding from the European Unions Horizon 2020 research and innovation programme under grant agreement NFFA-Europe Pilot Transnational Access Facility (Grant No. 101007417), having benefited from the access provided by ICN2 in Barcelona within the framework of the NFFA proposals [ID-900]. Finally, EEM appreciates Edital PRPPG 010/2024 Programa de Apoio a Jovens Professores(as)/Pesquisadores(as) Doutores(as) - JOVEMPESQ Project 24460.

\bibliography{refs}

\newpage
TOC graphic

\begin{figure}
\includegraphics[scale=0.8]{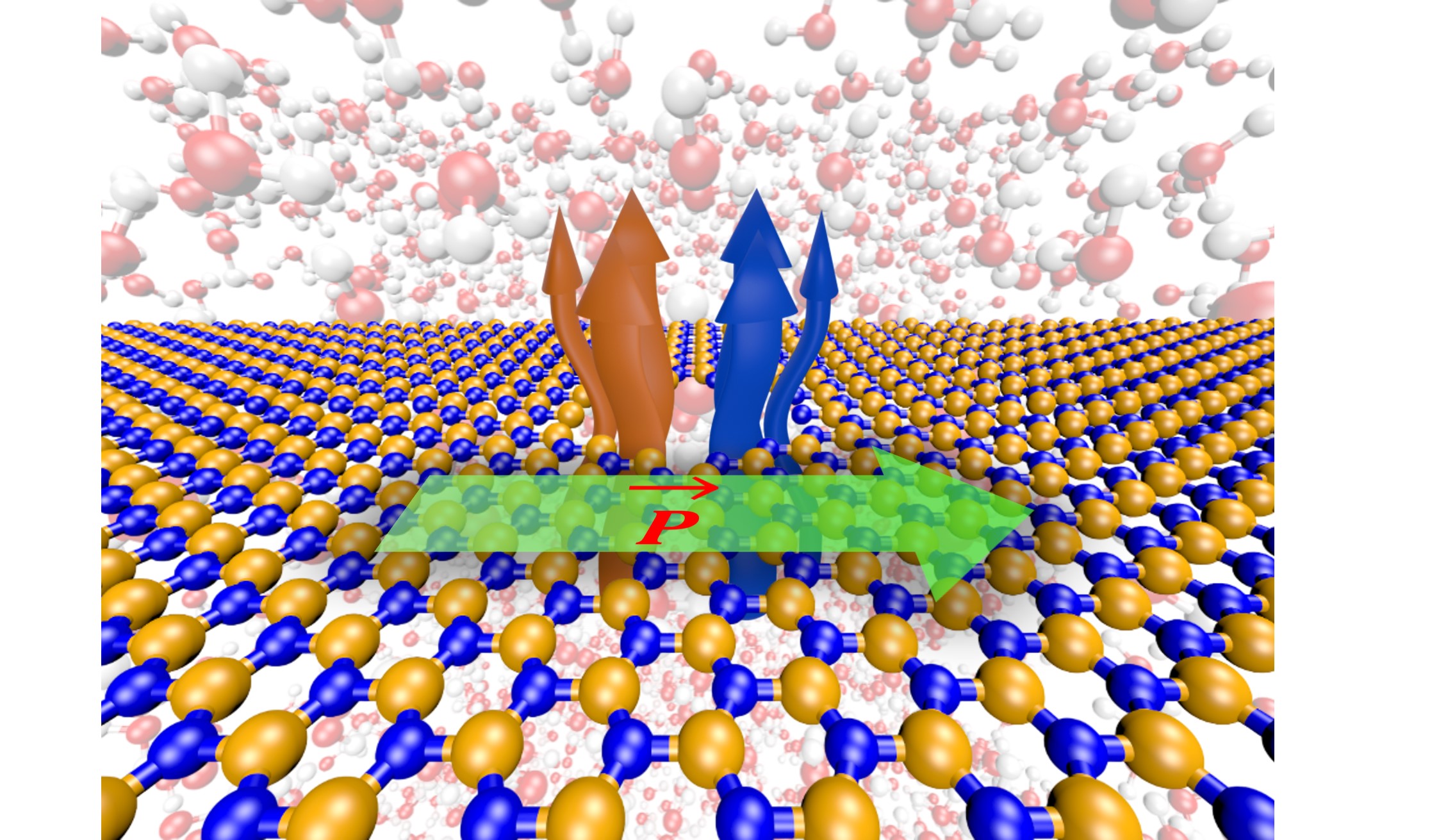}

\end{figure}

\end{document}